
\input phyzzx
\hoffset=0.2truein
\voffset=0.1truein
\hsize=6truein
\def\TITLEPAGE{\frontpagetrue}
\def\CALT#1{\hbox to \hsize{\tenpoint \baselineskip=12pt
        \hfil \vtop{
        \hbox{\strut CALT-68-#1}
        \hbox{\strut DOE RESEARCH AND}
        \hbox{\strut DEVELOPMENT REPORT}}}}
\def\CALTECH{
        \address{California Institute of Technology,
    Pasadena, CA 91125}}
\def\TITLE#1{\vskip.5in \centerline{\fourteenpoint#1}}
\def\AUTHOR#1{\vskip.2in \centerline{#1}}
\def\ABSTRACT#1{\vskip.2in \vfil \centerline
            {\twelvepoint \bf Abstract}
                     #1 \vfil}
\def\ENDTITLEPAGE{\vfil \eject \pageno=1}
\hfuzz=5pt
\tolerance=10000
\TITLEPAGE
\CALT{1875}
\TITLE{Exact Wavefunctions for non-Abelian Chern-Simons
Particles\footnote\dagger{This work supported in part
by the U.S. Department of Energy under Grant No.
DE-FG03-92-ER40701}}
\AUTHOR{Hoi-Kwong Lo\footnote\S{HKL@THEORY3.CALTECH.EDU}}
\CALTECH
\ABSTRACT{Exact wavefunctions for $N$ non-Abelian Chern-Simons (NACS)
particles are obtained by the ladder operator approach. The same method
has previously been applied to construct exact wavefunctions for
multi-anyon systems. The two distinct base states of the NACS
particles that we use are multi-valued and are defined in terms of
path ordered line integrals. Only strings of operators that preserve
the monodromy properties of these base states are allowed to act on
them to generate new states.}
\ENDTITLEPAGE
\eject

\def\Oab{\Omega_{\alpha \beta}}
\def\mab{m_{\alpha \beta}}
\def\nab{n_{\alpha \beta}}
\def\za{z_{\alpha}}
\def\zb{z_{\beta}}
\def\zab{z_{\alpha \beta}}
\def\zbar{\bar z}
\def\zbara{\bar z_{\alpha}}
\def\zbarb{\bar z_{\beta}}

\def\para{\partial_{\alpha}}

\def\parbar{\bar {\partial}}
\def\parbara{\bar {\partial}_{\alpha}}
\def\parbarb{\bar {\partial}_{\beta}}
\def\suma{\sum_{\alpha =1}^N}
\def\sumba{\sum_{\beta \neq \alpha}}
\def\sumalb{\sum_{\alpha < \beta}}
\def\prodalb{\prod_{\alpha < \beta}}
\def\aa{a_{\alpha}}
\def\ba{b_{\alpha}}
\def\adaga{a_{\alpha}^{\dag}}
\def\bdaga{b_{\alpha}^{\dag}}
\def\adagb{a_{\beta}^{\dag}}
\def\bdagb{b_{\beta}^{\dag}}
\def\psia{\hat \psi_I^{(0)}}
\def\psib{\hat \psi_{II}^{(0)}}
\chapter{Introduction}
It is now well-known that anyons---particles with arbitrary statistics
can exist in 2+1 dimensions.\Ref\rLei{M.~Leinaas and J.~Myrheim,
Nuovo Cimento B {\bf 37} (1977) 1; G.~A.~Goldin,R. Menikoff, and
D.~H.~Sharp, J. Math. Phys. {\bf 22} (1981) 1664; F.~Wilczek,
Phys. Rev. Lett. {\bf 48} (1982) 1144; {\bf 49} (1982) 957;
Y.~S.~Wu, Phys. Rev. Lett. {\bf 52} (1984) 2103.}
Owing to the topological gauge potential,
even non-interacting two-anyon states are not (symmetrized) tensor
products of single anyon states.\Ref\Arovas{D. P. Arovas, R. Schrieffer,
F. Wilczek and A. Zee, Nucl. Phys. B {\bf 251} (1985) 117.}
Thus the quantum mechanics of systems
of many anyons presents a challenge to field theorists.\Ref\Sen{D. Sen,
Phys. Rev. D {\bf 46} (1992) 1846.}
In general, the center of mass motion
of the system is not sensitive to the statistics and can be factored out.
For two anyons, the relative coordinates present us with a one-body problem
which can be solved in many cases. For $ N$ anyons, there are $N-1$ relative
coordinates, whereas there are $N(N-1)/2$ pairs of particles. These two
numbers match only when $N=2$. For $N>2$, the various pair-separation
coordinates (in terms of which the statistical gauge potential is easy to
write down) are not independent of each other. For this reason, not
a single three-anyon system has been completely solved.
The many-anyon system has been studied in the mean field approach.
\Ref\Fetter{A. L. Fetter, C. B. Hanna and R. B. Laughlin, Phys. Rev. B
{\bf 39} (1989) 9679; Y.-H. Chen, F. Wilczek, E. Witten and B. I. Halperin,
Int. J. Mod. Phys. B {\bf 3} (1989) 1001.}
Other methods that have been employed include the semi-classical
approximation\Ref\Ill{F. Illuminati, Phys. Lett. A {\bf 161} (1992) 323;
Phys. Rev. A {\bf 47} (1993) 3437. },
numerical studies\Ref\Sporre{M. Sporre, J. J. M. Verbaarschot and I. Zahed,
Phys. Rev. Lett. {\bf 67} (1991) 1813; SUNY-NTG-91-40; M. V. N. Murthy,
J. Law, M. Brack and R. K. Bhaduri, Phys. Rev. Lett. {\bf 67} (1991) 1817.},
perturbative analysis from the bosonic or fermionic
ends\Ref\per{C. Chou, Phys. Rev. D {\bf 44} (1991) 2533; D {\bf 45} (1992)
1433; C. Chou, L. Hua and G. Amelino-Camelia, Phys. Lett. B {\bf 286}
(1992) 329; D. Sen, Nucl. Phys. B {\bf 360} (1991) 397; R. Chitra and
D. Sen, Phys. Rev. B {\bf 46} (1992) 10923.},
and the most interesting of all,
the ladder operator approach: In the last couple of years, a
substantial subset of the exact multi-anyon wavefunctions in
a magnetic field has been found with this systematic analysis.
\Ref\mag{G. Dunne, A. Lerda, S. Sciuto and C. A. Trugenberger,
Nucl. Phys. B {\bf 370} (1992) 601; K. H. Cho and C. Rim,
Annals of Physics {\bf 213} (1992) 295.}
This soon generalized to free anyons\Ref\free{K. H. Cho, C. Rim
and D. S. Soh, Phys. Lett. A {\bf 164} (1992) 65.}
and anyons of multi-species.\Ref\multi{K. H. Cho, C. Rim and D. S. Soh,
preprint SNUTP-92-96.}
While a lot of states are still missing,
all the states in the lowest Landau level are obtained by this method.

In this paper, we apply the ladder operator approach to non-Abelian
Chern-Simons particles which may be regarded
as a generalisation of anyons.\Ref\CS{E. Witten, Comm. Math. Phys.
{\bf 121} (1989) 351;
E. Guadagnini, M. Martellini and
M. Mintchev, Nucl. Phys. B {\bf 336} (1990) 581; A. P. Balachandran,
M. Bourdeau and S. Jo, Int. J. Mod. Phys. A {\bf 5} (1990) 2423;
S. Elitzur, G. Moore, A. Schwimmer and N. Seiberg, Nucl. Phys. B {\bf 326}
(1989) 108-134.}
In the gauge that the Hamiltonian is
a free Hamiltonian, the wavefunctions (which have more than one component)
are multivalued with non-trivial monodromy properties given by a
monodromy matrix.\Ref\Lee{T. Lee and P. Oh, preprint SNUTP-93-4.}
By introducing statistical gauge potentials,
one has the liberty to work with single-valued wavefunctions.
However, since we find multivalued wavefunctions convenient to work with,
we will stick to them in the rest of this paper.

In section 2, we review the ladder operator formalism as applied to anyons.
This helps to highlight the differences between the cases of anyons
and non-Abelian C-S particles, the object under study in section 3.
In particular,
of the operators used for anyons, only a
subclass of operators, which preserve the monodromy
properties of the wavefunctions, are allowed to act on the C-S particles.
Nonetheless, our wavefunctions do cover the lowest Landau level.
As an application of our formalism, we compute the second virial coefficient
of NACS particles.
The same set of ladder operators apply to free NACS particles
with minor modifications.
We also consider systems of multi-species NACS particles.
Finally, the relevance of our work to
systems of vortices of finite gauge
groups is also discussed.

\chapter{Anyons}

The Hamiltonian for $N$ anyons with charge $e$ and mass $m$ moving on
a plane with a constant magnetic field $B$ (perpendicular to the plane)
is given by
$$H=\suma -{1 \over 2m} (\nabla_{\alpha} -i {\bf a}_{\alpha}-
ie{\bf A})^2,\eqno(1)$$
where the external gauge field $A^i=-{1\over 2}B \epsilon^{ij} x^j $
in the symmetric gauge and the statistical gauge potential
$$a^i_\alpha ({\bf x}_1, \dots,{\bf x}_N)=\nu \sumba
\epsilon^{ij}{x^j_\alpha -x^j_\beta \over |{\bf x}_\alpha -{\bf x}_\beta|^2}.
\eqno(2)$$

By a singular gauge transformation, we can remove ${\bf a}_\alpha$ from
the Hamiltonian at the expense of using multi-valued wavefunction
$$
\psi_{new} ({\bf x}_1,\dots,{\bf x}_N)=\exp \biggl(i \nu \sumalb
 \theta_{\alpha \beta} \biggr) \psi_{old} ({\bf x}_1,\dots,
{\bf x}_N).
\eqno(3)
$$
Using the complex notation $z=x^1+ix^2, \zbar=x^1-ix^2, \partial=
{\partial \over \partial z},\parbar={\partial \over \partial \bar z} ,$
the gauge transformed Hamiltonian becomes\refmark{\mag}
$$
H=\suma \biggl(-{2 \over m} \parbara \para + {e^2 B^2 \over 8m}
|\za|^2 \biggr) -{eB \over 2m}J, \eqno(4)
$$
where $J$ is the angular momentum operator in the singular gauge
$$
J=\suma (\za \para -\zbara \parbara). \eqno(5)
$$
Its eigenvalues are shifted from those in the symmetric gauge
by a constant ${1 \over
2} \nu N ( N-1).$ (See (10.a).)
It is convenient to extract a factor $\exp(-{1 \over 4}eB \suma
|\za|^2)$ from the wavefunction. Then the eigenvalue problem becomes
$$\hat H \hat \psi =(E-{1 \over 2}N \omega) \hat \psi ,\eqno(6.a)$$
$$J \hat \psi =j \hat \psi ,\eqno(6.b)$$
(with $\omega \equiv {eB \over m}$)
where the new Hamiltonian $\hat H$ and wavefunctions $\hat \psi$
are defined by
$$\hat H= \suma \biggl(- {2 \over m} \parbara \para +
{eB \over m} \zbara \parbara \biggr) , \eqno(7.a)$$
$$\hat \psi =exp \biggl({eB \over 4} \suma |\za|^2 \biggr) \psi.
\eqno(7.b)$$
Note that the ground state energy is shifted by ${1\over 2}N\omega.$
We impose two physical requirements for the wavefunctions. Firstly,
they must vanish at points of coincidences if $\nu \neq 0$ due to the
centrifugal potentials (hard-core requirement).
Secondly, they form Abelian representations of
the braid group.

Now we introduce the operators
$$
\adaga =\zbara-{2 \over eB} \para, \aa=\parbara , \eqno(8.a)$$
$$
\bdaga =\za -{2 \over eB} \parbara, \ba=\para  ,\eqno(8.b)$$
which satisfy $[\aa , \adagb]=[\ba , \bdagb]=\delta_{\alpha \beta}$,
all other commutators being zero.
With respect to these operators, the Hamiltonian $\hat H$ in (7.a) and the
angular momentum $J$ in (7.b) can be rewritten as
$$
\hat H=\omega \suma \adaga \aa, \eqno(9.a)$$
$$J=\suma ( \bdaga \ba -\adaga \aa). \eqno(9.b)$$

It is trivial to construct two distinct base states
(for $0 \leq \nu <2 $) with energy and
angular momentum eigenvalues:
$$\eqalign{\psia  &= \prodalb (\za-\zb)^{\nu},\cr
           E_I^0  &= {1 \over 2} N \omega,\cr
           j_I^0  &= {1 \over 2} \nu N(N-1),\cr} \eqno(10.a.)$$
$$\eqalign{\psib    &= \prodalb (\zbara- \zbarb)^{2-\nu},\cr
           E_{II}^0 &={1 \over 2} N \omega+{2-\nu \over 2}N(N-1)\omega,\cr
           j_{II}^0 &=-{2-\nu \over 2} N(N-1).\cr} \eqno(10.b)$$
The general strategy of the ladder operator approach is to construct
multi-anyon wavefunctions by acting with step operators on the base states
in (10.a) and (10.b).
We must, however, respect the statistics and hard-core requirements for the
resulting wavefunctions. In order to respect the statistics, we use only
symmetric combinations of the step operators. Consider the symmetric
operators
$$
C_{ln}= \suma {\adaga}^l {\bdaga}^n , \eqno(11)$$
where $l,n$ are non-negative integers such that $l+n \le N$.
(These operators form a basis in the ring of symmetric polynomials
in $2N$ variables.) They
are step operators in energy and angular momentum
which respect the statistics properties of the base states
$$
\eqalignno{[\hat H,C_{ln}]&=\omega l C_{ln}, &(12.a)\cr
           [J, C_{ln}]&=(n-l)C_{ln}. &(12.b)\cr}$$
They may, however, produce singular states
(states with non-vanishing wavefunctions at points of
coincidences) which have to be excluded
by hand.
We must identify which particular $C_{ln}$ produce
regular states. These operators can be safely applied to the base states.
Consider $C_{0n}$ first. With  (8.b), we have
$$
C_{0n}\psia =\bigl( \suma \za^n \bigr) \psia. \eqno(13)$$ Thus
they can be safely applied. For $C_{1m}$,
we have
$$
\eqalign{C_{1m}\psia  &=\suma (\zbara -\para) (\za -\parbara)^m \psia \cr
                      &=\suma (\zbara -\para) \za^m \psia \cr
                      &=-\suma \za^m \para \psia +
                \suma (\zbara \za^m -m \za^{m-1} ) \psia, \cr} \eqno(14)$$
where we set $eB=2$ for simplicity. The seemingly singular first term
is in fact regular because
$$\suma \za^m \para \psia=\suma \za^m\sumba {\nu \over \za - \zb} \psia
=\nu \sumalb {\za^m -\zb^m \over \za -\zb} \psia .\eqno(15)$$
By a similar proof, one can apply a sequence of operators of the form $C_{0n}$
followed by a sequence of operators of the form $C_{1m}$ to $\psia$
without generating any singularities. Moreover, states of the form
$$\psi_I^{(l)}=\prodalb (\zab)^{\nu +2l} , l=1,2,\dots \eqno(16)$$
are obtained from the action of $C_{on}$ on $\psia$. We can apply
a string of operators $C_{n_1 m_1} C_{n_2 m_2} \dots C_{n_i  m_i}$
with $\sum_{j=1}^i n_j \leq 2l$ to $\psi_I^{(1)}$ without generating
singularities, because such a string contains at most derivatives of order
$\sum_{j=1}^i n_j$ with respect to $\za$.

Thus we see that under suitable conditions, the step operators $C_{ln}$
can be safely applied to the base state $\psia$ to generate
regular new wavefunctions.
A similar analysis holds for the other base state $\psib$. Furthermore,
closed-form eigenfunctions generated by the action of combinations
of operators $C_{11},C_{10}$ and $C_{0m}$ only have been found, and they
can be expressed in terms of the Laguerre
functions.\refmark{\mag}
In particular, they do not involve the operators $C_{1m}$
with $m>1$, which however are allowed to act on $\psia$ to produce
regular wavefunctions.

One should also note that the step operator approach only generates a subset of
the whole spectrum of wavefunctions.\refmark{\mag}
If we naively set $\nu $ to be zero
or one, we obtain only a subset of the bosonic and fermionic wavefunctions.
Unlike the states generated by the step operators, the energies of
the missing states show non-linear dependence on the statistical parameter
$\nu$ in recent numerical studies.\refmark{\Sporre}
However, this is unimportant
for what follows.

\chapter{Non-Abelian Chern-Simons Particles}
Recently, there has been much interest in the non-Abelian generalization
of anyons. Non-Abelian Chern-Simons (NACS) particles carry non-Abelian
charges and interact with each other through the non-Abelian Chern-Simons
term. It has been argued that they may have applications in the fractional
quantum Hall effect.\Ref\nonabelion{G. Moore and N. Read, Nucl. Phys. B
{\bf 360} (1991) 362; B. Blok and X. G. Wen, Nucl. Phys. B {\bf 374}
(1992) 615.}
Consider a system of N particles each of which carries a statistical charge
corresponding to a representation $R_{l_\alpha} , \alpha=1, \dots ,N$ of
a non-Abelian gauge group, which for definite we take to be $G=SU(2)$.
In the holomorphic gauge, the dynamics of N free
$SU(2)$ NACS
particles is governed by the Hamiltonian\Ref\Ver{E. Verlinde, in {\it
Modern Quantum Field Theory}, Bombay International Colloquium 1990
(World Scientific, Singapore, 1991).}\refmark{\Lee}
$$
\eqalign{\hat H&=-\suma {1 \over m_{\alpha}} ( \nabla_{\zbara} \nabla_{\za}+
                     \nabla_{\za} \nabla_{\zbara}),\cr
   \nabla_{\za}&={\partial \over \partial \za}+{2 \over k} \sumba
                  {T_{\alpha}^a T_{\beta}^a \over \za -\zb},\cr
\nabla_{\zbara}&={\partial \over \partial \zbara},\cr} \eqno(17)$$
where k, a positive integer, is a parameter of the theory and $T_{\alpha}^a $
are the $SU(2)$-generators in the representation $R_{l_\alpha}$.

The wavefunctions take values in the tensor product of these representations.
$$\Psi \in R_{l_1} \otimes \dots \otimes R_{l_N}. \eqno(18)$$
We expand the single-valued wavefunction ${\bf \psi}$ in terms of the conformal
blocks ${\it F}_i \in R_{l_1} \otimes \dots \otimes R_{l_N}$
(which satisfy $\nabla_{\alpha} {\it F}_i=0$):
$$\Psi=\sum_i \psi_i {\it F}_i.  \eqno(19)$$
The Hamiltonian
acting on the new wavefunction is just the free Hamiltonian. However,
the complexity of the problem is hidden in the multivaluedness of
the wavefunctions $\psi_i$. (In fact, it is more ``natural'' to work with
the multi-valued wavefunctions $\psi_i$ than the original single-valued
wavefunctions ${\bf \psi},$ partly because
in the holomorphic gauge the Hamiltonian is not
hermitian with respect to the usual inner product. Instead, the inner
product is defined in the singular gauge and transformed back to the
holomorphic gauge by a non-unitary transformation function which has
to be taken into account in the definition of the inner product.
\refmark{\Lee})

{}From now on, we stick to the singular gauge. Consider $N$ NACS particles
in the same irreducible representation $R_l$ of $SU(2)$ moving in a
uniform external magnetic field $B$. We introduce operators $\aa,
\adaga,\ba,\bdaga$ as in eqn.(8) of section 2 and find that the
Hamiltonian is again given by eqn.(9). The only difference lies in
the constraints of the monodromy properties of the wavefunctions.
In the case of anyons, the wavefunctions have only one component and
monodromy leads to acquisition of phases, whereas NACS particles
have multi-component wavefunctions whose monodromy properties are
given by matrices.

We define
$$\Oab \equiv {2 \over k} \sum_a T_{\alpha}^a T_{\beta}^a .\eqno(20)$$
Note that $\sumalb \Oab$, $J$ and $\hat H$ commute with each other
and are thus good quantum numbers. ($J- \sumalb \Oab$ is the
angular momentum operator in the holomorphic gauge.) We will discuss
the diagonalization of $\sumalb \Oab$ later. For the time being,
let us assume this has been done and let $\psi_{I} \in R_{l_1}\otimes
R_{l_2}\otimes \dots \otimes R_{l_N}$ be a (position-independent)
eigenvector of $\sumalb \Oab$ with eigenvalue $\Omega$.
In analogy with
the anyon case, we propose applying the same ladder operator approach
with the following base states which are expressed as path-ordered
line integrals:\Ref\Kohno{T. Kohno, Ann. Inst. Fourier, Grenobie
{\bf 37} (1987) 139; in {\it Braids, Contemporary Mathematics} {\bf 78}
(1988).}
$$
\eqalignno{\psia(z_1, \dots, z_N)&=
           P \exp \biggl( \int_{\Gamma} \sumalb (\Oab -2 m_{\alpha
        \beta}I)
                  d \log(\za- \zb) \biggr) \psi_I , &(21)\cr
          \psib(z_1, \dots, z_N)&=
           P \exp \biggl( \int_{\Gamma} \sumalb (2n_{\alpha \beta} I-
                 \Omega_{\alpha \beta}) d \log(\zbara -\zbarb) \biggr)
                         \psi_I ,
                 &(22)\cr} $$
where $\Gamma$ is a path in the $N$-dimensional complex space with one
end point fixed and the other being $\zeta =(z_1, \dots, z_N).$
The $\mab$ ($\nab$)
depend on $\psi_I$ and are the
maximal (minimal) integers which make the wavefunctions
non-singular at the points of coincidences.
This is analogous to the requirement $0 \leq \nu <2$ in the anyon case.
Modulo the terms involving the identity matrix,
the first integrand is just the
flat Knizhnik-Zamolodchikov connection\Ref\KZ{V. G.
Knizhnik and A. B. Zamolodchikov, Nucl. Phys. B {\bf 247} (1984) 83.}
whereas the second is related to its antiholomorphic analogue.
One can easily check that these base states
have the desirable monodromy properties.
{}From (5),(6) and (7.a), we have
$$
\eqalign{\hat H \psia&=0,\cr
            E_I^0    &={1 \over 2} N \omega ,\cr
               J\psia&=\suma \za \sumba {(\Oab-2 \mab I)
                        \over \za -\zb}\psia\cr
                     &=\sumalb (\Oab-2 \mab) \psia\cr
                     &=(\Omega-2 \sumalb \mab )\psia, \cr
  \sumalb \Oab  \psia&=\Omega \psia,\cr
	} \eqno(23)$$
and
$$\eqalign{\hat H \psib&=\suma {eB \over m} \zbara \sumba
                 {(2 \nab I-\Oab) \over \zbara - \zbarb }\psib \cr
              &= \sumalb {eB \over m}(2 \nab I-\Oab)\psib\cr
              &={eB \over m} [2 \sumalb \nab-\Omega]\psib\cr}$$
$$    \eqalign{J\psib  &=- \suma \zbara \sumba {(2 \nab I- \Oab)
                             \over \zbara-\zbarb}
                         \psib \cr
                       &= \sumalb (-2\nab+ \Oab )\psib\cr
                       &=[-2 \sumalb \nab +\Omega] \psib,\cr
  \sumalb \Oab \psib   &=\Omega \psib, \cr} \eqno(24)$$
where we have used the relation\refmark\Kohno
$$[\sumalb \Oab, \Omega_{\gamma \delta}]=0 ,\eqno(25)$$
and the fact
that $\psi_I$ is an eigenstate of the operator $\sumalb \Oab$.
Mathematically, these
commutator relations are just consequences of the integrability condition
(infinitesimal pure braid relations) satisfied by the connection.
Physically, they follow from the fact that $\Omega$ is related to the angular
momentum $J$ which is invariant upon monodromy.

Now that we have found the analogous base states, we will apply the ladder
operators to them to generate new states. As before, the new states have to
respect the statistics. (The NACS particles in the same irreducible
representation are regarded as indistinguishable.)\Ref\scat{H.-K. Lo
and J. Preskill, preprint CALT-68-1867.}
Thus, we may only use
symmetric combinations of step operators. Also, we have to check that
the wavefunctions produced are regular at the point of coincidence.
There is, however, one crucial difference between the cases of anyons
and NACS particles. Even with symmetric step operators, there is no
guarantee that the monodromy properties of the wavefunctions are
preserved. Any combination of step operators which do not preserve
the monodromy properties of the wavefunctions are to be rejected.

First of all, let us consider $C_{0n}$. As before, we get
$$
C_{0n}\psia= \biggl( \suma \za^n \biggr) \psia . \eqno(26)$$
This shows that $C_{0n}$ can be safely applied to $\psia$ without
changing its monodromy properties or producing singularities.
Next we consider $C_{1m}$.
$$
\eqalign{C_{1m} \psia&=\suma (\zbara - \para) \za^m \psia \cr
                     &=-\suma \za^m \para \psia +
           \suma (\zbara \za^m -m \za^{m-1} ) \psia \cr
                     &=-\sumalb (\Oab-2\mab I) {\za^m -\zb^m \over \za - \zb}
                      \psia+ \suma ( \zbara \za^m -m \za^{m-1})\psia. \cr}
  \eqno(27)$$
For $m=0$,
$$
C_{10} \psia = \suma  \zbara \psia ,\eqno(28)$$
which clearly preserves the monodromy property of $\psia$.
When $m=1$, we have
$$
\eqalign{C_{11} \psia&=-\sumalb (\Oab-2\mab I) \psia + \suma (\zbara \za
                  -1) \psia \cr
                     &=[-\Omega+ 2\sumalb \mab +\suma(\zbara \za-1)] \psia.\cr}
 \eqno(29)$$
This shows that $C_{11}$ can be safely applied to the base state. We can say
more: strings made up of combinations of the operators $C_{0n}$ (n=1,2,...),
$C_{10}$ and $C_{11}$ act on the base state to generate physical states.
On the other hand, the operators $C_{1m}$ with $m>1$ and $C_{nm}$ with $n>1$
generally change
the monodromy property of base state. There is no obvious way of
constructing an admissible combination of operators involving them which
would preserve the monodromy property of the base state. We therefore
reject them as being unphysical and restrict the admissible set of
operators to those generated by $C_{0n}$,$C_{10}$ and $C_{11}$.

The crucial reason why the argument for $C_{1m}$ (with
$m>1$) and $C_{nm}$ (with $n>1$) as physical operators for anyons do
not carry over to the case of non-Abelian C-S particles is that the
the various monodromy matrices do not commute. In other words,
$$[\Oab, \Omega_{\gamma \delta}]\neq 0. \eqno(30)$$

As in the anyon case, there are
again missing states in the spectrum. However,
our wavefunctions do cover the entire lowest Landau level as
they involve the operators $C_{0n}$ only.

We now consider the construction of closed-form eigenfunctions. In the
case of anyons, for $P(z_1, \dots,z_N,\bar z_1,\dots,\bar z_N)
\prodalb (\za -\zb)^{\nu}$ to be an eigenfunction of $\hat H$,
it follows that the function $P$ has to satisfy
a modified differential equation.
$$
\suma \biggl(-{2 \over m} \para +{eB \over m} \zbara \biggr) \parbara
P -{2 \over m} \sumalb \biggl( {\parbara -\parbarb \over \za -\zb}
\biggr) \nu P=\biggl(E-\omega {N \over 2} \biggr)P \eqno(31)$$
An ansatz has been made to construct closed-form eigenfunctions.
All the solutions constructed can be expressed in terms
of the Laguerre Polynomials. They are generated by strings of operators
$C_{0n}$,$C_{10}$ and$C_{11}$ only. For NACS
particles, we get a similar equation for $ P$, but with $\nu$ replaced
by $\Oab-2\mab$. Nevertheless, since these operators are chosen to preserve
the monodromy properties of the states. There is every reason to believe
that the construction of closed-form solutions will go through
with ${1 \over 2} N(N-1) \nu$ replaced by $\Omega-2 \sumalb \mab$.

Finally, we come to the diagonalization of $\sumalb \Oab$. Consider
the identity
$$
\sum_{a} ( T_1^a + T_2^a + \dots + T_N^a )( T_1^a +T_2^a+ \dots +T_N^a)
=\suma T^a_{\alpha}T^a_{\alpha} +2 \sumalb T^a_{\alpha} T^a_{\beta}.
\eqno(32)
$$
For $SU(2)$, the left-hand side gives the Casimir operator $J(J+1)$ of
the ``spin'' of the composite made up of the $N$ particles, and the first
term on the right-hand side gives the sum of the Casimir operators
$\suma J_{\alpha} (J_{\alpha}+1)$ of the ``spins'' for the individual
particles. (Here we abuse the word ``spin'' for the internal $SU(2)$
symmetry group. The physical spin (which is a scalar in 2+1 d)
of a NACS particle in the $j$
representation is given by
${2 \over k} J(J+1)$.
Thus, for $SU(2)$, we have
$$
\Omega \equiv \sumalb \Oab={1 \over k}[J(J+1)- \suma J_{\alpha}
(J_{\alpha}+1)].\eqno(33)$$

We just decompose the composite state into
irreducible representations and $\sumalb \Oab$
would be diagonal in that basis.
Actually, we can do better than that. It is easy to check that
the operator $T^z_1 +T^z_2+ \dots +T^z_N$ commutes with $\hat H$,
$J$ and $\sumalb \Oab$. Thus they can be simultaneously diagonalised.

\chapter{Second virial coefficient and the large k limit}
In this section, we compute the second virial coefficients for some
simple systems of NACS particles. To do so, we need to know
all the two-particle states only.

First of all, consider two identical NACS particles in the $j={1 \over 2}$
representation of $SU(2)$. From the addition rule for angular
momenta, we find that the resulting states
consist of a triplet with $\Omega={1 \over
2k}$ and a singlet with $\Omega=-{3 \over 2k}.$
For $N=2$, $\Omega$ plays the role of the anyon phase,
$\nu$. Let us recall the formula derived
by Arovas {\it et al.}\refmark{\Arovas} for the second virial
coefficient of anyons,
$$
B(\nu=2j+\delta,T)=\lambda^2_T (-{1 \over 4}
 + | \delta | -{1 \over 2} \delta^2),\eqno(34)
$$
where $|\delta| <2$. Note that it has a cusp at Bose values $\nu
=2j.$
By taking the average over the four two-body states, the second virial
coefficient of the NACS particles is given by
$$B(j={1 \over 2},T)=\lambda^2_T  [-{1 \over 4}+{3 \over 4k} -{3 \over
8 k^2}]. \eqno(35) $$

For two particles with $j=1$, the resulting states have ``spins''
$2$, $1$, and $0$ (with $\Omega = {2\over k}$,
$-{2 \over k}$, and $-{4 \over k}$ and degeneracies
$5$, $3$ and $1$ respectively).
We remark that all these states are bosonic if $k=1$. When $k=2$, the
singlet is a bosonic state whereas others are fermionic. For $k>4$,
all the states are anyonic with $|\nu| <1 $.
For $k>1$, we have
$$B(j=1,T)=\lambda^2_T [-{1 \over 4} +{20 \over 9 k}-{8 \over 3 k^2}].
\eqno(36)$$

Now we come to the large $k$ limit. For two particles belonging to
a representation $j$ with $\lim_{{j \to \infty}\atop {k \to \infty} }
{j^2 \over k}=a <1$, we
approximate the sum over all the resulting ``spins''
$r \le 2j$ by an integral.
For example, the $|\delta|$ term is given by
$$
\eqalign{&{1
 \over (2j+1)^2} \sum_{r=0}^{2j} (2r+1) {1 \over k}\bigl|[r(r+1)-
2j(j+1)] \bigr|\cr
\sim &{1 \over kj^2} \int_{r=0}^{\sqrt2 j} -r (r^2- 2j^2)\cr
=    &{j^2 \over k}, \cr} \eqno(37)$$
where in the second line we approximate the sum by an integral and
divide it into two parts (which happen to be equal) according to the
sign of $r^2 -2 j^2$. The $\delta^2$ term can be evaluated in a
similar manner.
Hence, we get
$$B=\lambda^2_T [-{1 \over 4}+{j^2 \over k}- { j^4 \over 3 k^2}]
   =\lambda^2_T [-{1 \over 4}+a-{a^2 \over 3}].
\eqno(38)$$

If $a \to 0$, the last term may be discarded
and the second virial coefficient of the NACS particle in the $j$
representation (with physical spin ${2 \over k} j(j+1)$)
is the same as that of an anyon with
{\it half} of the physical spin
as its statistical parameter.

\chapter{Concluding Remarks}
(1) The $N$-free-NACS-particle problem can be solved by a similar method.
The free Hamiltonian in the ``anyon'' gauge is given by
$$H=\suma -{2 \over m} \parbara \para . \eqno(39)$$
The subtlety is that our base states
become unnormalizable.\refmark{\free}
Let us define $r =(\suma |\za|^2)^{1 \over 2}$ and consider
$$
H M(r)\psia = \psia \times -{1\over 2m} [ d^2/dr^2 +(1/r)(2N-1 +2 \Omega
-4 \sumalb \mab)]
M(r). \eqno(40)$$
We have eigenfunctions of the form
$$
M_{\mu}(r)=r^{-\mu} J_{\mu} (kr),\eqno(41)$$
(where $\mu=2(N-1+\Omega-2 \sumalb \mab)$)
with eigenvalues $\hbar k^2/ 2m$ for $H$.

(2) Let us now consider the construction of $C_{lm}$ for multi-species
non-Abelian C-S particles (particles in various irreducible representations).
In this case,
when we construct $C_{0n}$, we do so for each irreducible
representation $R$ and symmetrize over
particles in this irreducible representation only.\refmark{\multi}
Let us call the resulting operator $C^R_{0n}$.
If we construct $C^R_{10}$ and $C^R_{11}$ in a similar manner, we
find that these operators have to be rejected: They do not preserve
the monodromy properties of the base states because
$[\Oab,\Omega_{\gamma \delta}] \ne 0,$ and we are no longer
summing over all the particles.
Therefore, for $C_{10}$ and $C_{11}$ we do sum over all the particles
in the various irreducible representations.

(3) Note that $C_{01}$ and $C_{10}$ represent
center of mass excitations. The operator $C_{10}$ was also analyzed by
Johnson and Canright,\Ref\Johnson{M. D. Johnson and G. S. Canright,
Phys. Rev. B {\bf 41} (1990) 6870.}
while $C_{11}$ is directly related to
the Lie group generator of $SU(1,1)$.\Ref\Awaji{T. Awaji and M. Hotta,
Int. J. Mod. Phys. B {\bf 6} (1992) 1229.}

(4) We remark that the operators of $C_{1m}$ and $C_{nm}$ ($m,n>2$) do
preserve the monodromy property of the base state, if $\psi_I$ in
eqns.(20) and (21) is chosen to be a simultaneous eigenstate of all
$\Oab$. The statistics is ``Abelianized'' in this case. This situation
occurs, for example, for some models of non-Abelian vortices of finite
gauge groups such as the quarternion group.

(5) The same ladder operator approach may well apply to non-abelian
vortices of finite gauge groups.\refmark{\scat}\Ref\finite{F. Wilczek
and Y.-S. Wu,
Phys. Rev. Lett. {\bf 65} (1990) 13; M. Bucher, Nucl. Phys. B {\bf 350}
(1991) 163; M. G. Alford, K. Benson, S. Coleman, J. March-Russell and
F. Wilczek, Phys. Rev. Lett. {\bf 64} (1990) 1632; Nucl. Phys. B
{\bf 349} (1991) 414; J. Preskill and L. Krauss, Nucl. Phys. B {\bf 341}
(1990) 50; F. A. Bais, P. van Driel and M. de Wild Propitius,
Phys. Lett. B {\bf 280} (1992) 63; Nucl. Phys. B {\bf 393} (1993)
547.}
Unfortunately, we generally do not know how to construct connections
which would produce the desirable monodromy in this case.

(6) In a recent paper,\Ref\eos{A. Dasnieres de Veigy and S. Ourvy,
``Equation of state of an anyon gas in a strong magnetic field,''
preprint IPNO/TH 93-16.}
Dasnieres de Veigy and Ourvy derived the equation
of state of an anyon gas in a strong magnetic field at low temperatures.
The idea is that at sufficiently low temperatures, excitations to higher
Landau levels can be neglected. Thus one may consider only the lowest Landau
states of the anyons, which are covered by the step operators.
In fact, apart from the statistical phase factor, the multi-anyon
states in the lowest Landau level are tensor-product states of
the individual anyon states. By regularizing the grand partition
function with a harmonic potential, the equation of state can be
obtained. The same decoupling principle should
apply to NACS particles. For a fixed base state, modulo the statistical
term involving $\Oab$, the multi-particle wavefunctions
in the lowest Landau level are
again tensor products of individual particle states. Therefore,
in principle, one should be able to derive the equation
of state of NACS particles in a strong magnetic field at low
temperatures.
\bigskip
We thank John Preskill and Piljin Yi for useful discussions.
\refout

\bye